\documentclass[useAMS,usenatbib]{mn2e}
\usepackage{graphicx,amssymb,color}
\usepackage[normalem]{ulem}

\newcommand{\rev}{  }

\title[Post-MS YORP Debris II]
{Post-main-sequence debris from rotation-induced YORP break-up of small bodies II: multiple fissions, internal strengths
and binary production}

\author[Veras \& Scheeres]{Dimitri Veras$^{1,2}$\thanks{E-mail: d.veras@warwick.ac.uk}\thanks{STFC Ernest Rutherford Fellow},
Daniel J. Scheeres$^3$
\\
$^{1}$Centre for Exoplanets and Habitability, University of Warwick, Coventry CV4 7AL, UK
\\
$^{2}$Department of Physics, University of Warwick, Coventry CV4 7AL, UK
\\
$^{3}$Department of Aerospace Engineering Sciences, The University of Colorado, Boulder, CO 80309, United States
}

\pubyear{2019}

\begin{document}
\label{firstpage}
\pagerange{\pageref{firstpage}--\pageref{lastpage}}
\maketitle

\begin{abstract}
Over one quarter of white dwarfs contain observable metallic debris from the breakup
of exo-asteroids. Understanding the physical and orbital history of this debris would enable
us to self-consistently link planetary system formation and fate. One major debris reservoir is generated
by YORP-induced rotational fission during the giant branch phases of stellar evolution, where
the stellar luminosity can exceed the Sun's by four orders of magnitude. Here, we determine the
efficacy of the giant branch YORP effect for asteroids with nonzero internal strength, and model post-fission evolution by 
imposing simple analytic fragmentation prescriptions. We find that even the highest realistic internal strengths
cannot prevent the widespread fragmentation of asteroids and the production of a debris field
over 100 au in size. We compute the number of successive fission events as they occur in 
progressively smaller time intervals as the star ascends the giant branches, providing a way to generate size distributions of
asteroid fragments. The results 
are highly insensitive to progenitor stellar mass.  We also conclude that the ease with which
giant branch YORP breakup can generate binary asteroid subsystems is strongly dependent on internal strength.
Formed binary subsystems in turn could be short-lived due to the resulting luminosity-enhanced BYORP effect. 
\end{abstract}

\begin{keywords}
Kuiper belt: general – minor planets, asteroids: general – planets and satellites: dynamical evolution and stability – stars: AGB and post-AGB – stars: evolution – white dwarfs.
\end{keywords}

\section{Introduction}

Understanding the full lifecycle of planetary systems requires connecting their formation
around main sequence stars with their fate around white dwarfs. The glue that binds these two
endpoints are giant branch stars, which arguably produce the most violent changes to the system evolution.
Such violence is appropriately showcased by how intact planets dominate observations 
of main-sequence exo-planetary systems whereas shattered asteroidal debris dominate observations of white
dwarf exo-planetary systems.

Asteroids are hence a key component of our understanding. White dwarfs represent stars
which are both dense enough to stratify all infalling material into its constituent atoms 
\citep{schatzman1958} and differentiated enough to feature a photosphere consisting of entirely
hydrogen and/or helium. Consequently, observations of infalling material are unambiguous and direct,
and represent the most substantial available windows into the bulk chemical composition of exo-planetary material.
Between one-quarter and one-half of white dwarfs feature rocky debris 
\citep{zucetal2003,zucetal2010,koeetal2014} and 20 different metals have now been observed
in these remnants \citep[e.g.][]{gaeetal2012,juryou2014,melduf2017,xuetal2017,haretal2018,holetal2018,doyetal2019,swaetal2019}.

Asteroids are also showcased in these systems in other respects. Minor planets have been observed orbiting
the white dwarfs WD 1145+017 \citep{vanetal2015} and SDSS J1228+1040 \citep{manetal2019}
in real time\footnote{These are the only known individual exo-asteroids, as main-sequence analogues 
have not yet been discovered.} and might generate the transit dips seen in ZTF J0139+5245 \citep{vanetal2019}. 
Asteroids are also the most likely progenitors for the 
over 40 known white dwarf
discs \citep[e.g.][]{zucbec1987,gaeetal2006,farihi2016,denetal2018}. In fact, the minor planet orbiting SDSS J1228+1040 is 
embedded within the debris disc, and resides well within the star's Roche radius, suggesting that the asteroid has high internal strength
\citep{veretal2017}. That object is probably better characterised as an asteroid-sized ferrous 
planetary core fragment, which also has not been observed in main-sequence systems. 

The dynamical pathways leading to asteroidal signatures in white dwarf planetary systems remain
a subject of debate and suffer from a dearth of dedicated studies \citep{veras2016a}. Although these pathways
are first shaped during planet formation and subsequent main-sequence evolution, during the giant
branch phase the system is transformed in three major ways: 

\begin{itemize}

\item {\bf Stellar mass loss.}
A giant branch star will lose between about one-half and three-quarters of its mass {\rev non-constantly} with 
time. For Solar-mass stars, most mass is lost at the tips of the red giant branch and asymptotic giant
branch phases; for more massive stars, the red giant branch phase is effectively suppressed. In general,
stellar mass loss changes all of the orbital elements of surrounding material 
\citep{veretal2013a,doskal2016a,doskal2016b}. However, imposition of the reasonable assumption of isotropy
for the mass loss results in simplified orbital changes, when the only variations are in the semimajor axis,  
eccentricity, and argument of pericentre \citep{omarov1962,hadjidemetriou1963}. A further simplification occurs when considering
orbital changes within about several hundred au: here the so-called adiabatic approximation can be employed
\citep{veretal2011}, such that (effectively) only the semimajor axis changes, and at a rate proportional to the mass loss.
Consequently, unless one considers the effect of mass loss on Oort clouds 
\citep{alcetal1986,paralc1998,verwya2012,veretal2014a,veretal2014b,stoetal2015,caihey2017}, 
most planets and asteroids will simply increase their semimajor
axes by a factor of 2-4 due to giant branch mass loss.

\item {\bf Stellar radius expansion}
Upon ascending the giant branches, stars will increase their radii by a factor of hundreds; in the
solar system, at least Mercury and Venus will be swallowed \citep{schcon2008,veras2016b}. The resulting
tidal interaction with planets are significant and have been well-studied 
\citep{villiv2009,kunitetal2011,musvilla2012,adablo2013,nordspi2013,viletal2014,
madetal2016,staetal2016,galetal2017,raoetal2018,sunetal2018}. In contrast, giant branch tidal effects on asteroids have not
been studied in any appreciable way, although asteroids are small enough to not extend the critical tidal
engulfment distance beyond the stellar surface. 

\item {\bf Stellar luminosity enhancement}
The greatest danger to asteroids is stellar radiation. Stars at the tip of the asymptotic giant branch are 
tens of thousands of times more luminous than the sun. The consequences for planetary atmospheres and surfaces
remain largely unexplored {\rev \citep{schetal2019}}, but should not alter their orbits and spins. However, the same is not
true for asteroids. Asteroids are easily flung about \citep{veretal2015a,veretal2019a} due to a radiative effect
known as the Yarkovsky effect \citep{radzievskii1954,peterson1976}. Asteroids can also easily be spun up
to breakup speed \citep{veretal2014c} (hereafter Paper I) through another radiative effect known as the YORP effect
{\rev \citep{rubincam2000}}. The result is both an orbital rearrangement and a debris field spreading out to
hundreds of au. 

\end{itemize}

These asteroids and this debris are then later propelled towards the white dwarf through a combination
of Poynting-Robertson drag \citep{stoetal2015,veretal2015b} and gravitational interactions with 
planets \citep{bonetal2011,debetal2012,frehan2014,veretal2016,antver2016,antver2019,musetal2018,smaetal2018} or moons
\citep{payetal2016,payetal2017} or binary stellar companions \citep{bonver2015,hampor2016,petmun2017,steetal2017,steetal2018}. These substellar bodies may then break up en route towards the white dwarf \citep{makver2019} or within its Roche radius, forming a debris disc {\rev \citep{graetal1990,jura2003,debetal2012,beasok2013,veretal2014d,malper2020a,malper2020b}} or becoming embedded in an existing disc \citep{griver2019}. 

Hence, an understanding of the distribution and type of debris in giant branch systems crucially allows one to link them with observations of planetary debris in white dwarf planetary systems. The need for this understanding will become more urgent with {\rev discoveries of major planets orbiting white dwarfs \citep{ganetal2019}} and the expected order-of-magnitude increase of {\rev debris-rich} systems due to the entire known population of white dwarfs increasing by a factor of 8 in the year 2018 \citep{genetal2019}.

Despite this urgency, the only paper which has so-far addressed giant branch YORP break-up of asteroids is 
Paper I. That paper introduced the concept of ubiquitous destruction due to giant-branch YORP
effects, but modelled the process with the most basic assumptions: with no internal strength nor fragmentation prescription.
This paper aims to generalise Paper I by incorporating these two features, a task which is facilitated by the
recent investigation of \cite{scheeres2018}. Our paper does not merge the YORP and Yarkovsky effects in 
a self-consistent framework, which even for the simplest YORP formalism would be beyond the scope of this study
\citep{veretal2015a}. Our paper also does not strive to achieve the level of fine detail often required to model the solar system 
YORP effect with features like local topography and avalanches \citep[e.g.][]{golkru2012,staetal2014,yuetal2018}.

We aim instead to provide useful order-of-magnitude values which can be used in future studies.
In Section 2 we perform our analysis. We discuss our results 
in Section 3 and conclude in Section 4.

\section{Asteroid evolution due to YORP}

\subsection{Spin-up}

We begin by quantifying the extent to which an asteroid spins up over time
due to giant branch luminosity. We can express this evolution from
\cite{scheeres2007} and Paper I, but in a similar form to equation
1 of \cite{scheeres2018}, as

\begin{equation}
\frac{d\omega(t)}{dt} =
\frac{3 \mathcal{C} \Phi}{4 \pi \rho R_{0}^2 a(t)^2 \sqrt{1-e^2}}
\left(\frac{L_{\star}(t)}{L_{\odot}}\right)
.
\label{omegadot}
\end{equation}

\noindent{}Here $\omega$ is the asteroid's spin rate, $R_0$ its initial radius, $a(t)$
its semimajor axis, $\rho$ its density (taken to be 2 g/cm$^3$ throughout the paper) 
and $e$ its eccentricity (taken to be 0 throughout the paper). The luminosity of the star is $L_{\star}(t)$ and
the solar radiation constant $\Phi$ is taken to be $1 \times 10^{17}$ kg$\cdot$m/s$^2$.
The constant $\mathcal{C}$ expresses the amount of asymmetry and obliquity in the asteroid;
following \cite{scheeres2018}, we adopt the two values $\mathcal{C} = 10^{-2}$ and 
$\mathcal{C} = 10^{-3}$ as bounds on the shape. 

Time dependencies are explicitly stated 
for spin, luminosity and semimajor axis, because these are the only three values which we assume change 
continuously with time. The asteroid's eccentricity remains fixed because we employ the 
adiabatic approximation
for mass loss \citep{veretal2011} as the semimajor axes we will be considering are all under about 600 au.

When the asteroid undergoes fission, we will replace $R_0$ with a fissioned component.
Otherwise, $R_0$ remains fixed. In reality, $R_0$ could change continuously in time due to sublimation,
particularly if the asteroid contains an internal reservoir of volatiles 
\citep{jurxu2010,jurxu2012,faretal2013,radetal2015,malper2016,genetal2017,malper2017a,malper2017b}. Orbital changes could result from anisotropic outgassing {\rev \citep{maretal1973,froric1986,krolikowska2004,stejac2014,veretal2015c}} whereas spin changes could directly result from equation (\ref{omegadot}) due to a reduction of $R_0$.

Integration of equation (\ref{omegadot}) requires a value for the initial asteroid spin,
a stellar evolution profile, and an integration duration. To simplify our analysis and 
reduce the phase space to explore, we model only spin increases and
adopt two choices for $\omega(0)$. The first is $\omega(0) = 0$ rad/s, and
the second is $\omega(0) = -\sqrt{4 \pi G \rho/3}$. This latter choice is the negative
value of the breakup spin rate for a strengthless rubble pile \citep{scheeres2018},
and hence would require the greatest injection of positive spin to break apart.
The stellar evolution profiles are taken from the {\tt SSE}
code \citep{huretal2000}. These profiles include a Reimers mass loss prescription 
\citep{reimers1975,reimers1977} along the red giant branch phase and a superwind prescription
along the asymptotic giant branch phase \citep{vaswoo1993} with the default values
(including Solar metallicity) given within {\tt SSE}. We integrate for the entire red giant and asymptotic
giant branch phases, and then for 1 Gyr of white dwarf cooling. Doing so allows us 
to confirm that white dwarf radiation has little effect on asteroids whose orbits have already
been pushed outward by a factor of a few.

\begin{figure}
\includegraphics[width=9cm]{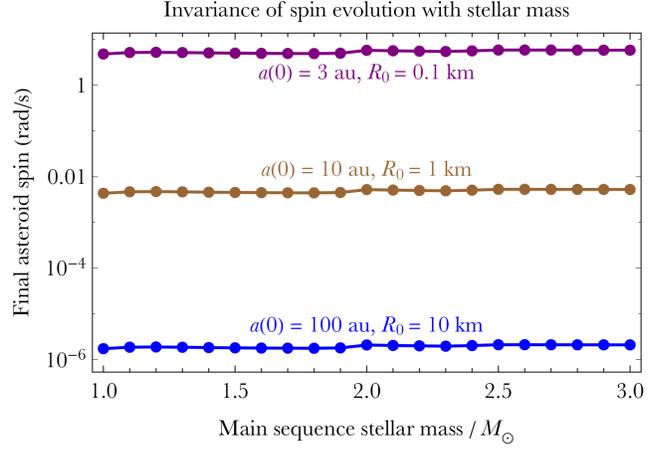}
\caption{
Demonstration that the final YORP-induced asteroid spin after giant branch evolution is effectively independent of the stellar mass, despite the different stellar evolutionary sequences and timescales. Every dot represents a separate integration. The evolutions here do not include break-up, which almost certainly would have occurred for the architectures represented in the upper two curves.
}
\label{steldep}
\end{figure}

One helpful finding is that the interplay between $a(t), L_{\star}(t)$ and the duration of the giant branch
phases {\rev produces} results which are independent of stellar mass in an order-of-magnitude manner,
removing that variable from further consideration.
To demonstrate this effect, in Fig. \ref{steldep}, we plot the final asteroid spin for three different initial sets of parameters, after
each was integrated separately around 31 different main-sequence stellar masses ranging from
$1.0M_{\odot}$ to $3.0M_{\odot}$.\footnote{This range bounds the progenitor masses
for the vast majority of known white dwarf planetary systems \citep{treetal2016}.} These parameter
sets straddle extremes (Set 1: $a(0)~=~3$~au and $R_0~=~100$~m; 
Set 2: $a(0)~=~10$~au and $R_0~=~1$~km;  
Set 3: $a(0)~=~100$~au and $R_0~=~10$~km) and all assume $\mathcal{C} = 10^{-3}$. We note that in the first two of the cases, 
the asteroids would have probably broken apart long before these final spins were achieved: this plot is 
just for demonstration purposes, to help show the invariance of the results with respect 
to stellar mass.

We henceforth assume a main-sequence progenitor mass of $2.0 M_{\odot}$ throughout the remainder of the paper.

\subsection{Internal strength}

We now compute how strong a rubble-pile asteroid must be to resist giant branch YORP break-up.
By ``asteroid strength'' we refer to the strength of the bonds between the rubble components (e.g. between different 
boulders and grains), which is usually less than 1 kPa \citep[Table 3 of][]{scheeres2018}. The individual boulders and 
grains themselves have much different and higher internal strengths, usually greater than 1 MPa 
\citep[Tables 1-2 of][]{schetal2015}. We will discuss
this dichotomy and the implications further in Section 3. For the remainder of Section 2, however,
we will assume that each successive fission produces fragments which are larger than the 
constituent boulders and grains.

The failure spin rate $\omega_{\rm fail}$ of an asteroid can be approximated as 
{\rev \citep{sansch2014,stejac2014,scheeres2018}}

\begin{equation}
\omega_{\rm fail}^2 \approx \frac{4\pi G \rho}{3} + \frac{2\sigma}{\rho R_{0}^2} \left(\frac{2}{3}\right)
,
\label{omegafail}
\end{equation}

\noindent{}where $\sigma$ is its
tensile, uniaxial strength. The fraction of $(2/3)$ in the rightmost term is a value arising from a representative
angle of friction. By equating the final spin state of an asteroid $\omega_{\rm final}$ with $\omega_{\rm fail}$,
we then obtain the critical tensile strength $\sigma_{\rm crit}$ which is the minimum that the asteroid would need to harbour in order
to survive post-main-sequence evolution:

\begin{figure*}
\centerline{
\includegraphics[width=9cm]{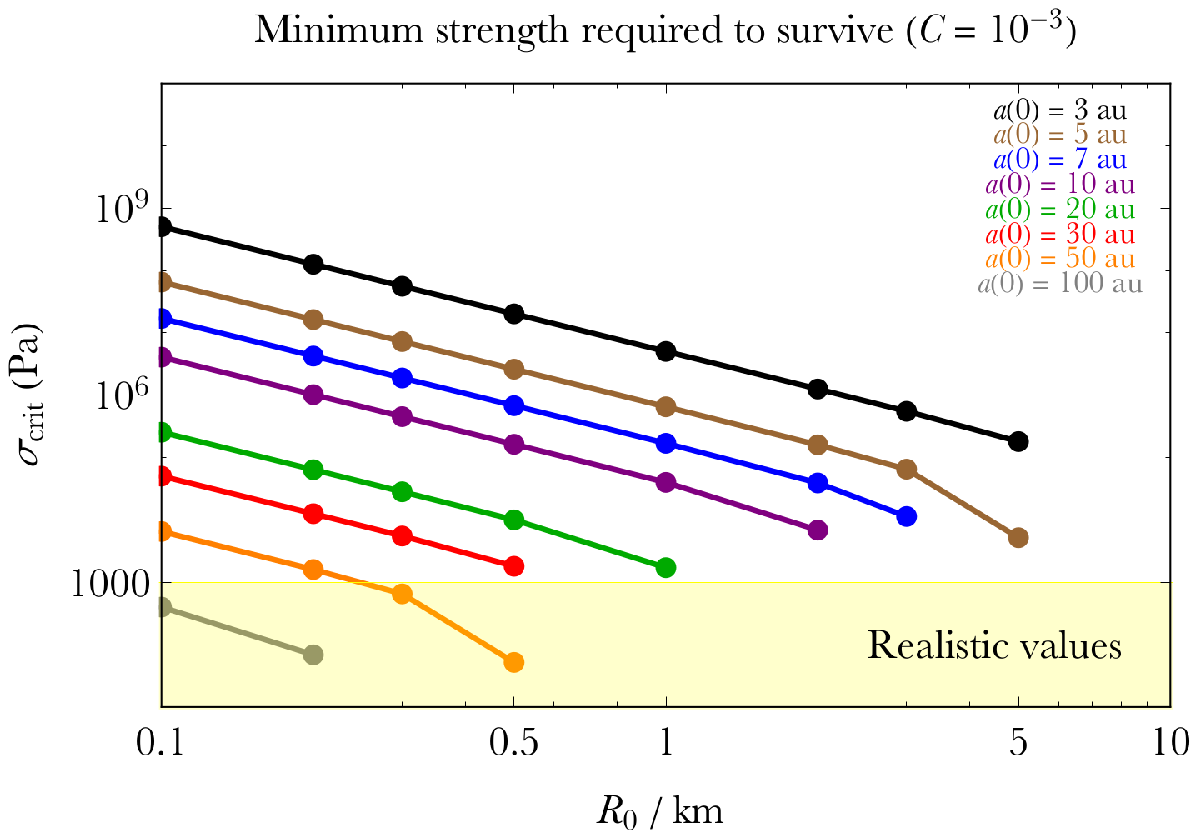}
\ \ \ \ \ \
\includegraphics[width=9cm]{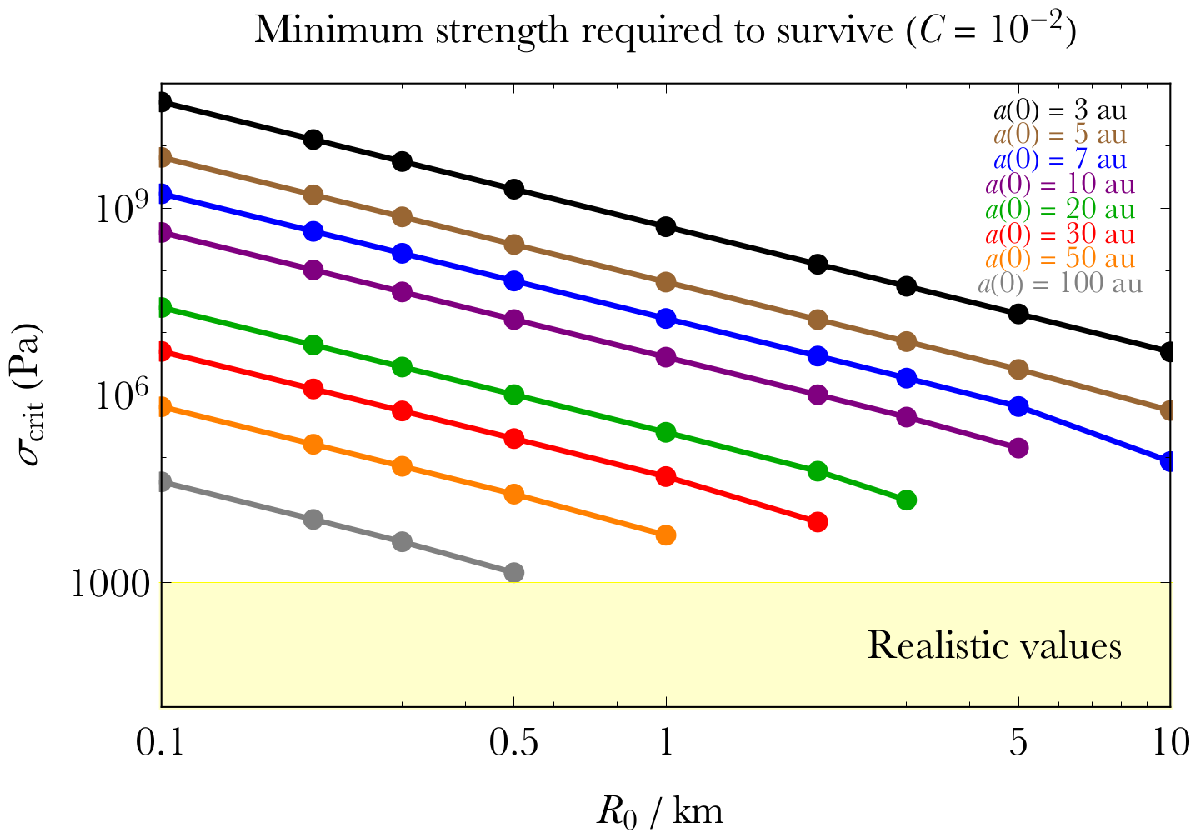}
}
\caption{
The critical (or minimum) internal strength an asteroid would need in order to survive giant branch YORP spin-up
{\rev for $a(0)$ values ranging from 3 au (top curves) to 100 au (bottom curves)}. The
left and right plots illustrate the shape- and obliquity-based bounding cases corresponding to $\mathcal{C} = 10^{-3}$ and $\mathcal{C} = 10^{-2}$.
A uniform grid of 72 integrations was performed on each plot in $R_0$---$a(0)$ space: each dot represents an integration 
for an asteroid that would be destroyed without any internal strength, and the rightmost dot on each curve represents the largest asteroid that is destroyed for a given $a(0)$. The plots demonstrate that asteroids require unrealistically high values of internal strength to survive rotational fission.
}
\label{sigmacrit}
\end{figure*}

\begin{equation}
\sigma_{\rm crit} = \frac{1}{4} R_{0}^2 \rho \left(3 \omega_{\rm final}^2 - 4 \pi G \rho \right)
.
\label{sigcrit}
\end{equation}

We compute $\sigma_{\rm crit}$ for a range of asteroid semimajor axis and radii in Fig. \ref{sigmacrit}.
On each plot, we performed 72 integrations (for eight different $a(0)$ values and nine different $R_0$ values). 
The right endpoint of each curve indicates the largest asteroid which experienced fission; larger asteroids,
which do not undergo fission, need not have any internal strength. Note that 
survival occurs only for the largest asteroids at the furthest distances from the star.

For the asteroids which do undergo fission, in almost every case, $\sigma_{\rm crit}$ is unrealistically high (see, for example, Table 3 of \citealt*{scheeres2018}
and, for meteorite strengths, Table 1 of \citealt*{schetal2015}). Hence, the inclusion of internal strength does not change
the results of Paper I. The only cases where a realistic internal strength can prevent breakup are for $a(0) \gtrsim 50$ au
and $R_0 \lesssim 0.5$ km.

\begin{figure*}
\includegraphics[width=18cm]{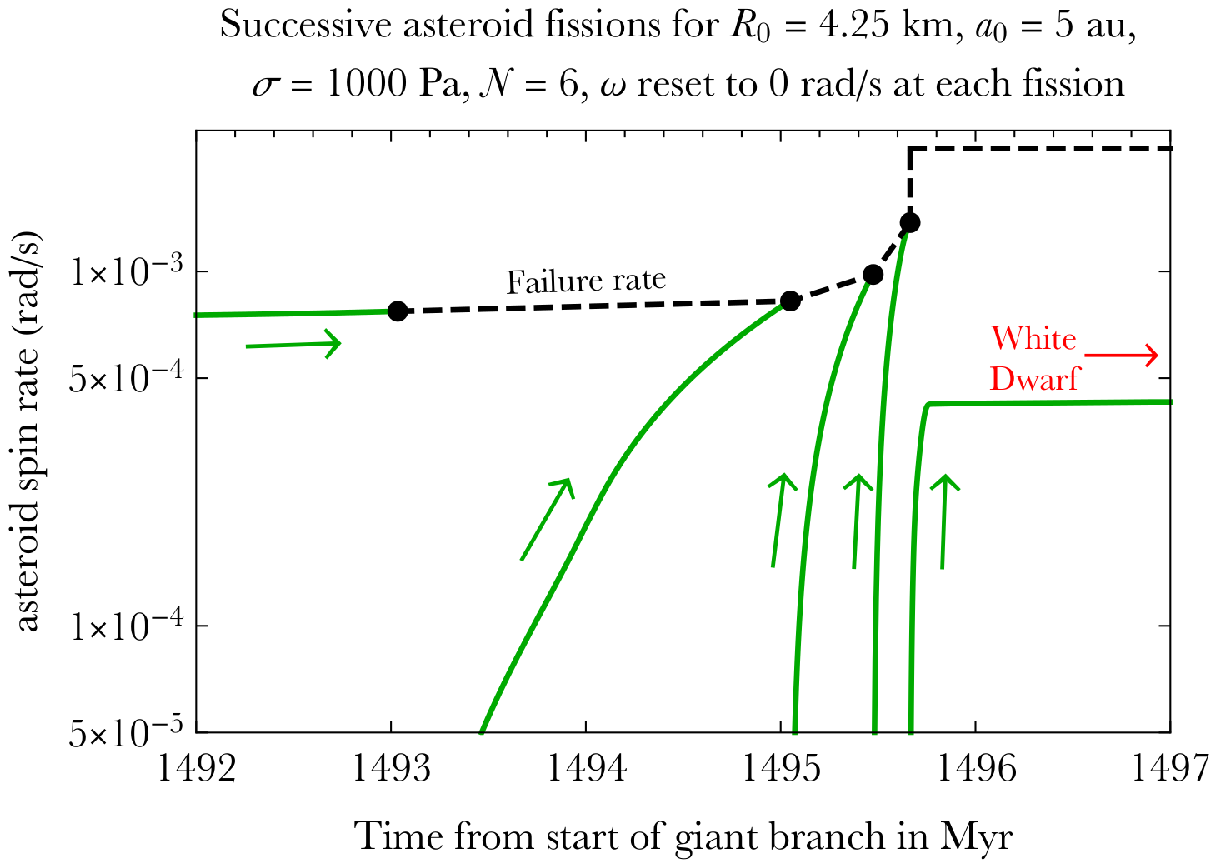}
\caption{
Spin evolution of a $R_0 = 4.25$ km asteroid which undergoes four fissions during giant branch evolution. In each fission, the progenitor splits into
6 equal child asteroids, generating a total of about 1300 asteroids of radii $R_4 = 0.39$ km. Because the asteroids have internal strength ($\sigma = 10^3$ Pa), the failure
spin rate increases with each fission. The star becomes a white dwarf soon after the last fission, effectively halting
any further spin rate increases.
}
\label{spinevo}
\end{figure*}

\begin{figure*}
\centerline{
\includegraphics[width=9cm]{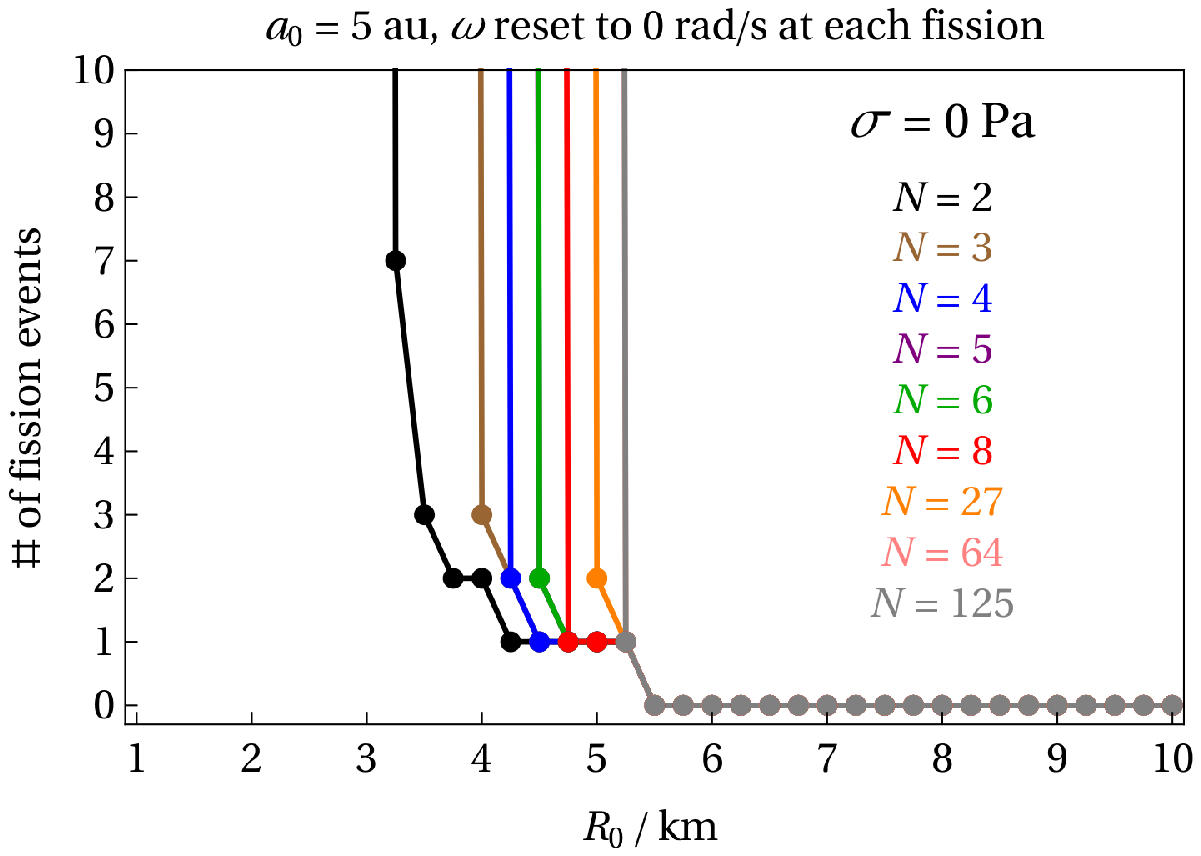}
\ \ \ \ \ \
\includegraphics[width=9cm]{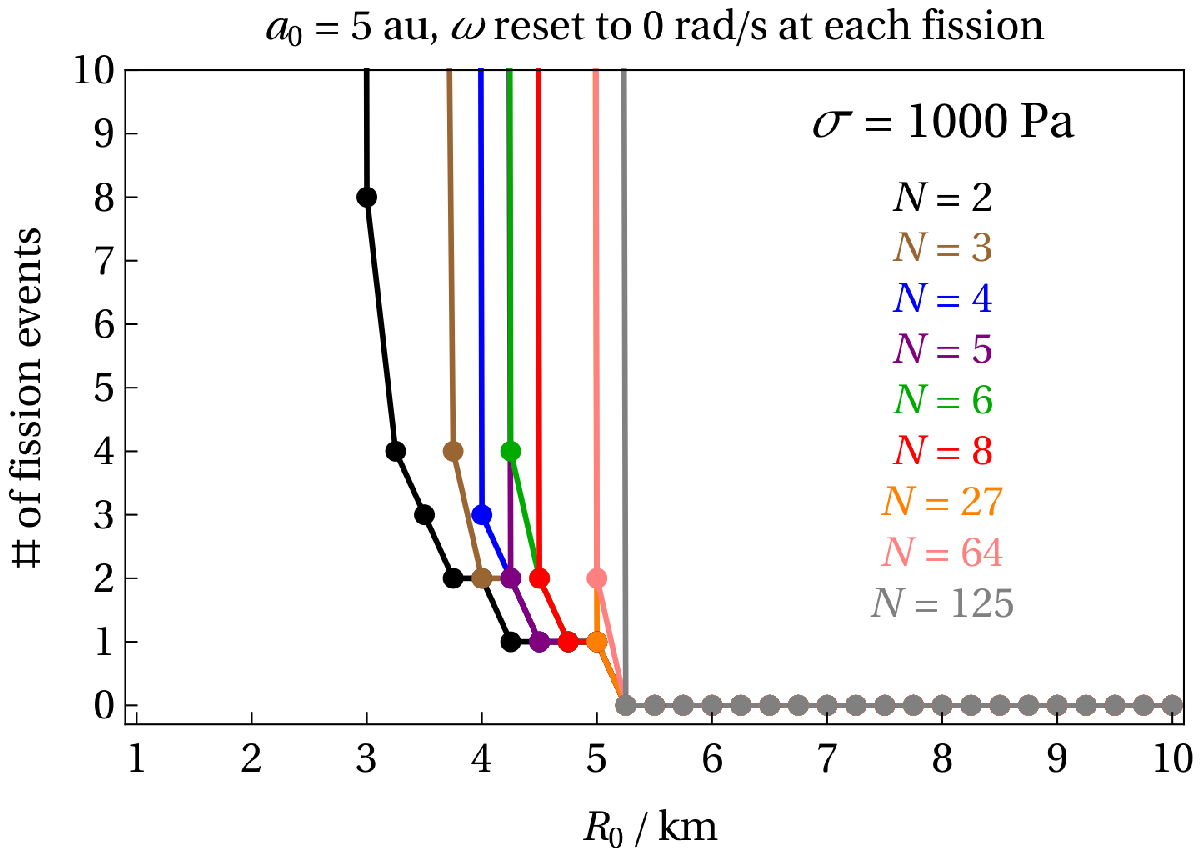}
}
\caption{
Number of fissions for $a(0) = 5$ au asteroids, both without strength (left panel)
and with high strength (right panel). After each fission, the spin rate is reset
to stationary. {\rev In both plots, the curves from left to right correspond to an increasing
number of fissions from 2 to 125.}
}
\label{a5om0plots}
\end{figure*}

\begin{figure*}
\centerline{
\includegraphics[width=9cm]{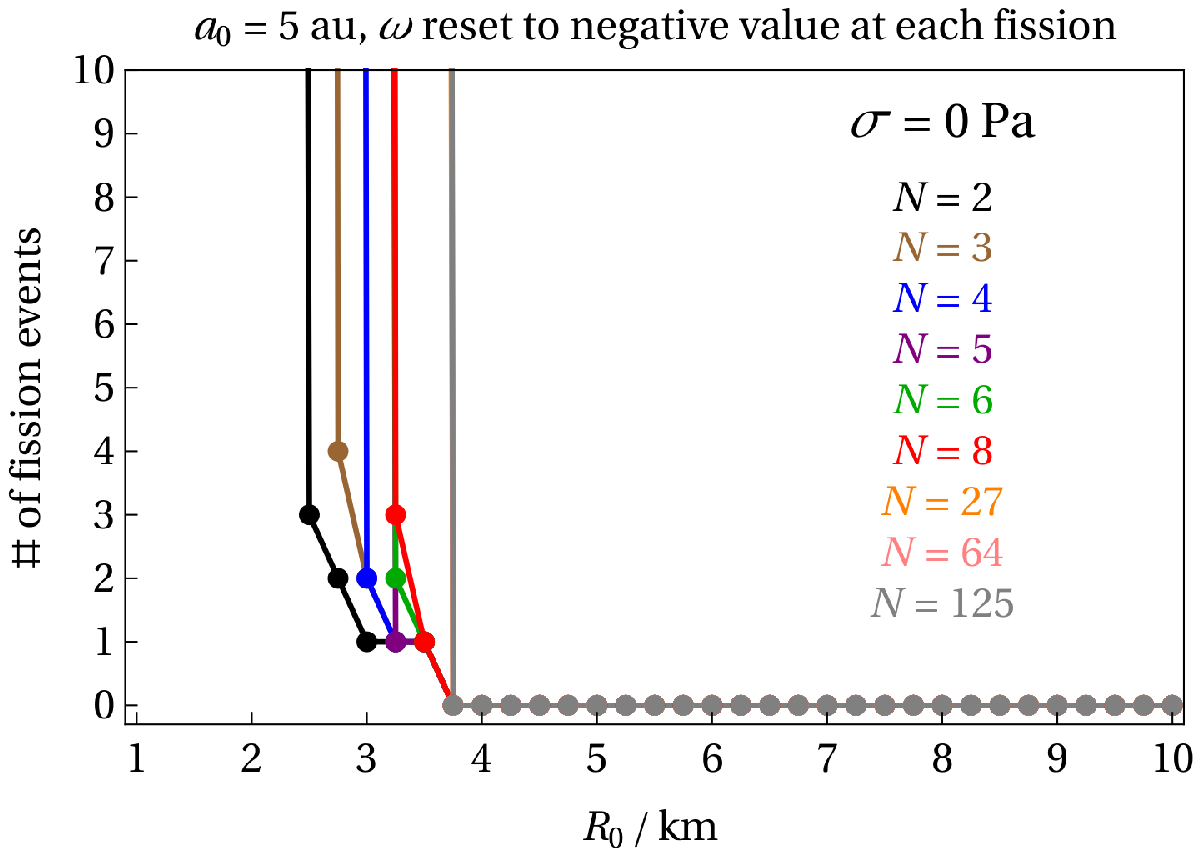}
\ \ \ \ \ \
\includegraphics[width=9cm]{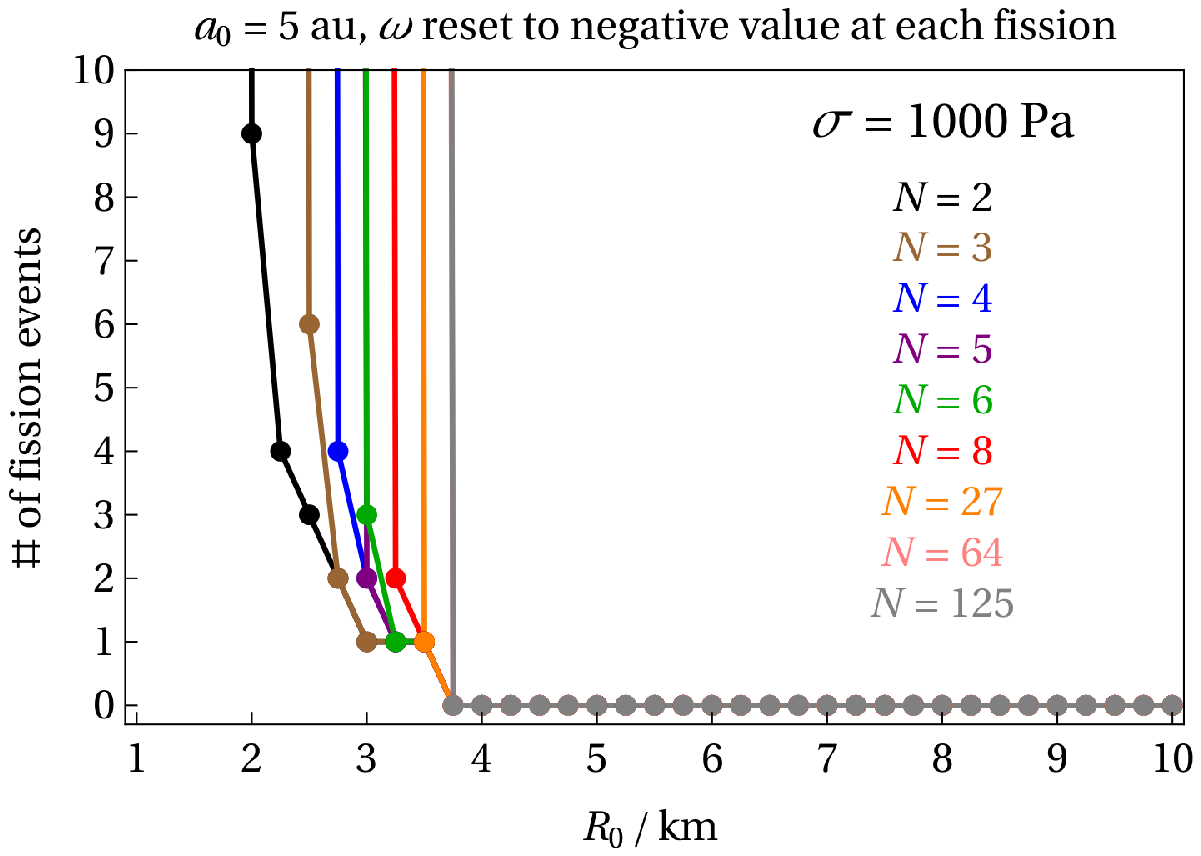}
}
\caption{
Same as Fig. \ref{a5om0plots}, except here after each fission the spin 
rate is reversed and set to the extreme value of the cohesionless limit.
}
\label{a5omm1plots}
\end{figure*}

\begin{figure*}
\centerline{
\includegraphics[width=9cm]{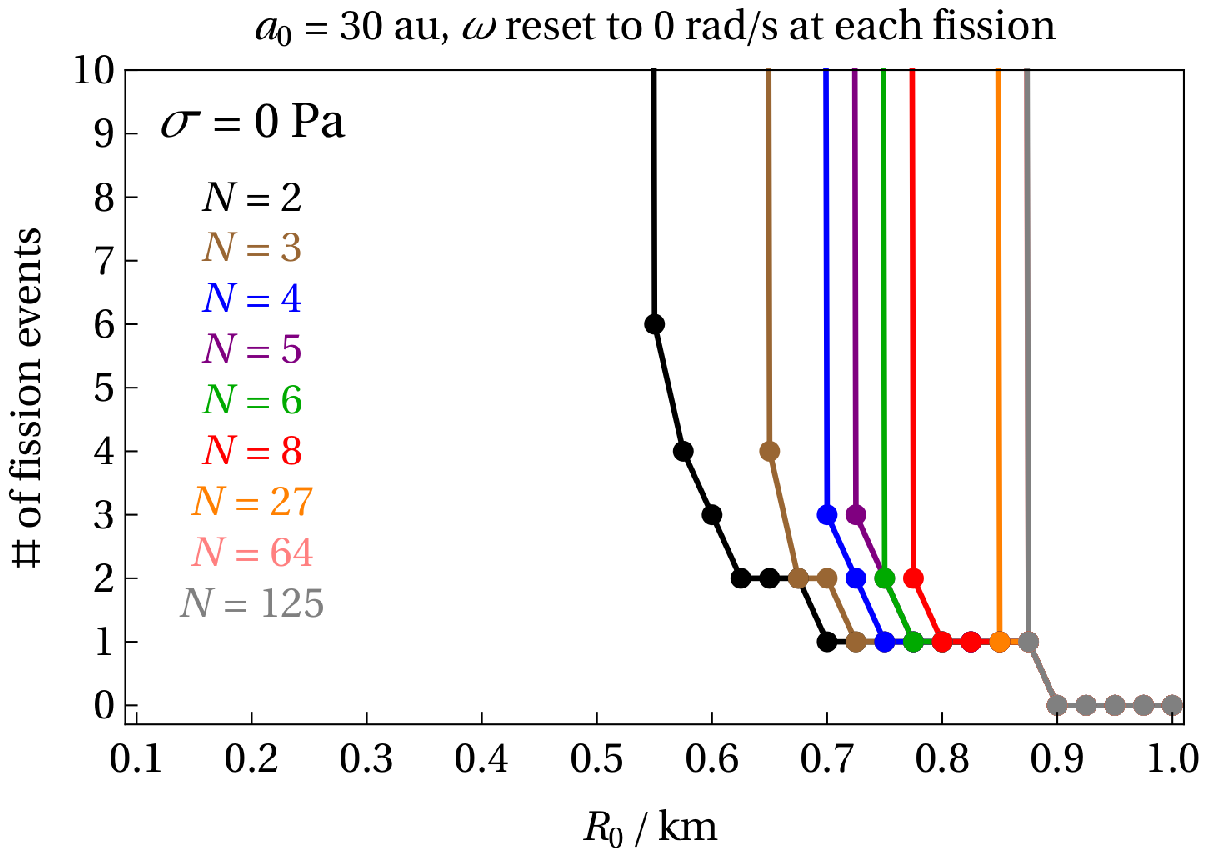}
\ \ \ \ \ \
\includegraphics[width=9cm]{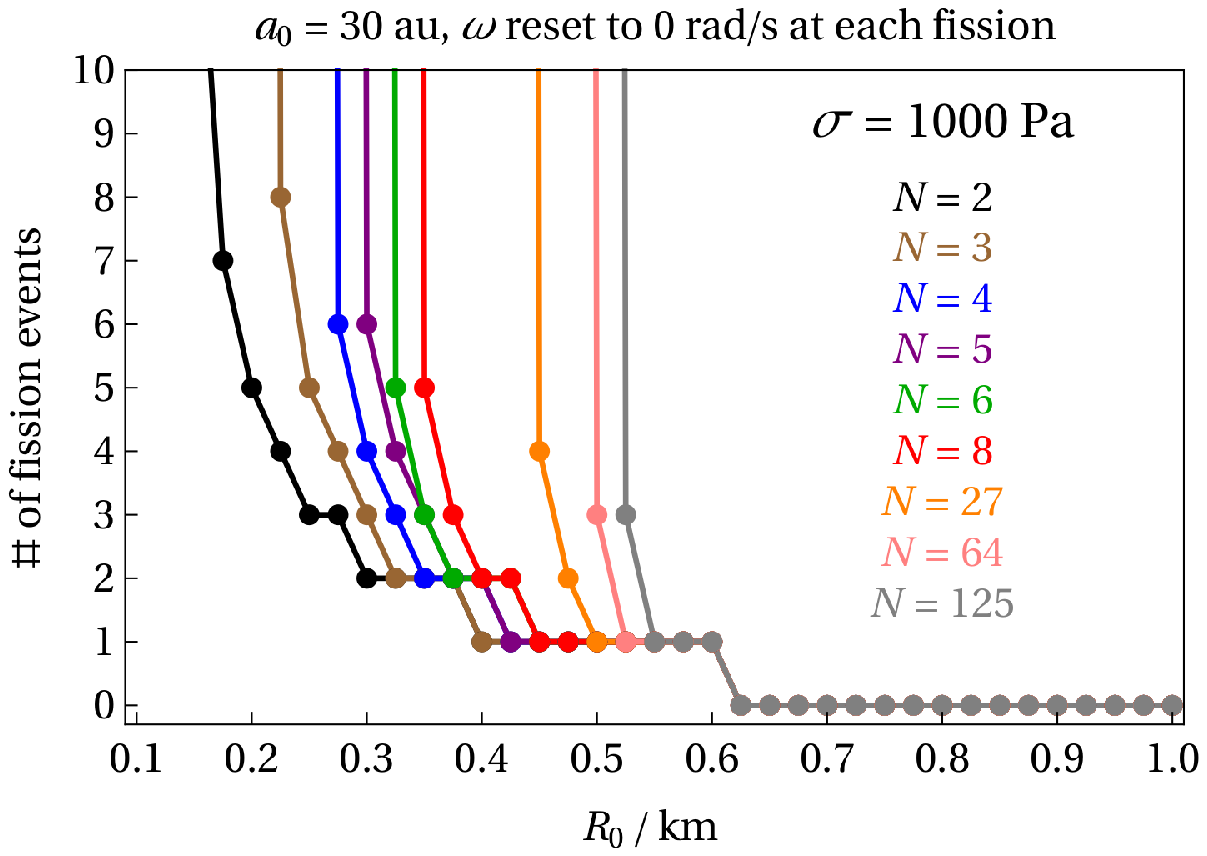}
}
\caption{
Number of fissions for $a(0) = 30$ au asteroids, both without strength (left panel)
and with high strength (right panel).
}
\label{a30plots}
\end{figure*}

\begin{figure}
\includegraphics[width=9cm]{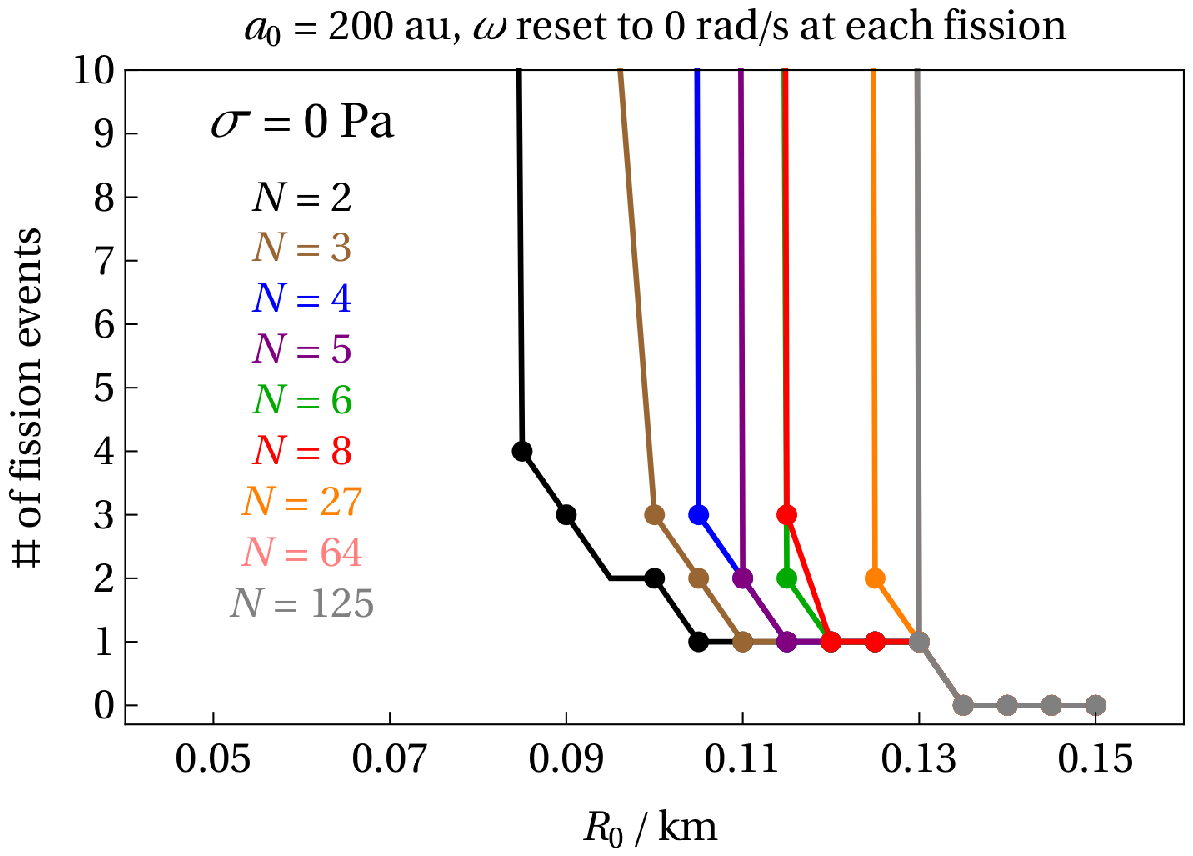}
\caption{
Number of fissions for $a(0) = 200$ au asteroids.
}
\label{a200plot}
\end{figure}

\subsection{Fragmentation}

Having established the robustness of the destructive giant branch YORP process,
we now investigate post-fission evolution, which was not quantified in Paper I.
\cite{scheeres2018} provided a straightforward fragmentation prescription that
is well-suited for our purposes. He assumed that the asteroid will split into $\mathcal{N}$ 
equal bodies of radii

\begin{equation}
R_1 = \mathcal{N}^{-1/3} R_0
,
\end{equation}

\noindent{}such that mass is conserved and after the $i$th fission, there will be a total of $\mathcal{N}^i$ bodies,
each with radii 

\begin{equation}
R_i = \mathcal{N}^{-i/3} R_0
.
\end{equation}

The reduction of asteroid radius due to a fission increases its spin rate of spin 
acceleration (equation \ref{omegadot}) and,
if the asteroid contains internal strength, the failure spin rate (equation \ref{omegafail}).
The asteroid spin rate increase is steeper than the failure spin rate
increase. Combined with the
steep increase in stellar luminosity, particularly along the tip of the asymptotic giant
branch phase, the result can be a relentless fission cascade. This cascade would cease
-- or at least our fragmentation model would no longer be applicable -- at a 
value of $R_i$ corresponding to the constitutents of the asteroid (boulders or grains).

The realistic maximum number of fissions depends largely on the chosen value of $\mathcal{N}$. For example,
$\mathcal{N} = 2$ and $i = 10$ reduces a 10~km asteroid into about $10^3$ separate 1~km asteroids,
whereas $\mathcal{N} = 5$ and $i = 10$ reduces a 10~km asteroid into about $10^7$ separate 50~m asteroids.

How the spins of the child asteroids are reset after a fission significantly affects the subsequent number of fissions, but
depends on details of the disruption dynamics 
that we do not model here. One reasonable 
possibility is that the post-fission spin rate is equal to the previous fission spin rate.
We consider two other extreme cases, one where the child asteroid spins reset to zero, and the other where their spin
rates are reversed to the rubble pile limit ($-\sqrt{4 \pi G \rho/3}$). 

In order to demonstrate a typical asteroid evolution with this fragmentation model, we have plotted
spin rate versus time for a $R_0 = 4.25$ km with high internal strength ($\sigma = 10^3$ Pa) in Fig. \ref{spinevo}. 
In this plot, we assume that the post-fission spin rate resets to zero,
and $\mathcal{N} = 6$.
The parent asteroid undergoes four fissions, ultimately producing about 1300 child asteroids, with 
successive radii of $R_1 = 2.3$ km, $R_2 = 1.3$ km, $R_3 = 0.70$ km and $R_4 = 0.39$ km.
The locations of the fissions are indicated by black dots, and their increase (due to a large finite
strength of $\sigma = 1000$ Pa combined with asteroid radius reduction; see equation \ref{omegafail}) 
is tracked with the black dashed lines.  The last horizontal black dashed segment is never reached.

In the figure, all of the fissions occur within 3 Myr. This brief destructive epoch occurs at the tip of the asymptotic
giant branch phase (for a $2.0 M_{\odot}$ progenitor star). The time to each subsequent fission becomes progressively shorter.
The rate of increase of the spin rate of the last child generation is reduced to a negligible value abruptly as the
star becomes a white dwarf. The white dwarf's rapid luminosity decrease \citep[e.g.][]{altetal2009} combined
with the expanded orbit of the child asteroids effectively flatlines their spin rate curves.

We next determine the number of fissions experienced for different types and locations of asteroids.
We report the results in a series of 7 plots in Figs. \ref{a5om0plots}-\ref{a200plot}. All plots illustrate
the number of fissions versus initial asteroid radius, for $\mathcal{C}~=~10^{-3}$. 
The entire range of asteroid radii
which were used in the integrations is shown; an absence of a dot indicates that the number of fissions
for that integration was greater than 10. A value of 0 fissions indicate
that the asteroid survived giant branch YORP spin-up. One common characteristic of all plots is that
there exists only a narrow range of radii for which the number of fissions is between 0 and 10.

In the first two figures (figure \ref{a5om0plots}-\ref{a5omm1plots}), we consider asteroids
with $a(0) = 5$ au. The only difference between these two figures is how the spin is reset after
each fission (to 0 in the first figure, and to the negative cohesionless limit in the second).
This difference changes the critical radii at which fission occurs by 1-2 km, and
has a much stronger dependence than the inclusion of internal strength (right panels).

In the next figure (figure \ref{a30plots}), we consider asteroids
with $a(0) = 30$ au. Consequently, the asteroid radii ($x$-axes) for which the number of fissions lies between 0 and
10 is lowered by an order of magnitude from figures \ref{a5om0plots}-\ref{a5omm1plots}. Also, the 
difference in the two plots in figure \ref{a30plots} demonstrates
that the inclusion of internal strength has a stronger relative effect than in figures \ref{a5om0plots}-\ref{a5omm1plots}. 

The final figure (figure \ref{a200plot}) illustrates a $a(0)~=~200$~au case, which might be considered extreme
in main-sequence systems, but not here. The plot illustrates that asteroids with radii on the order of 100 m
will regularly undergo fissions and hence produce debris in those regions of white dwarf planetary systems.

\subsection{Forming binary asteroids}

A fission can produce bound components in some circumstances, forming (in the case of 2 or 3 child asteroids)
a ``binary asteroid'' or ``ternary asteroid''.  Achieving this state requires the subsystem to attain
net positive energy. 

\cite{scheeres2018} {\rev provided} a criterion for escape for $\mathcal{N}~=~2$, which 
requires the child asteroids to be unequal in radii and mass (unlike in the last subsection). 
Now assume that the two children have radii $R_{1{\rm a}}$ and $R_{1{\rm b}}$. 
Further assume $R_{1{\rm a}} > R_{1{\rm b}}$ such
that 

\begin{equation}
R_{1{\rm a}} = k_1 R_{1{\rm b}}
\label{kratio}
\end{equation}

\noindent{}with $k_{1} > 1$. Then the parent and child radii are related through

\begin{equation}
R_0 = \left(R_{1{\rm a}}^3 + R_{1{\rm b}}^3\right)^{1/3},
\end{equation}

\begin{equation}
R_{1{\rm a}} = \frac{R_0 k_1}{\left(k_{1}^3 + 1\right)^{1/3}},
\end{equation}

\begin{equation}
R_{1{\rm b}} = \frac{R_0}{\left(k_{1}^3 + 1\right)^{1/3}}.
\end{equation}

\noindent{}The critical spin rate below which these two children would remain bound to each other as
a binary asteroid \citep{scheeres2018} can be written as

\begin{equation}
\omega_{\rm bound}^2 = \frac{8\pi G \rho}{3} \frac{R_{1{\rm a}}^3 + R_{1{\rm b}}^3}{\left(R_{1{\rm a}} + R_{1{\rm b}} \right)^3} 
= \frac{8\pi G \rho}{3} \frac{1 + k_{1}^3}{\left(1 + k_{1}\right)^3}
.
\end{equation}

The minimum value of $k_{1}$ for which the fissioned binary would remain bound 
($k_{1,{\rm min}}$) is then given
through $\omega_{\rm bound} = \omega_{\rm fail}$, yielding

\begin{equation}
\frac{1 + k_{1,{\rm min}}^3}{\left(1 + k_{1,{\rm min}}\right)^3}
=
\frac{1}{2} +
\frac{\sigma}{2 \pi G \rho^2 R_{0}^2}
.
\end{equation}

\noindent{}Solving for $k_{1,{\rm min}}$ gives

\begin{equation}
k_{1,{\rm min}} = 
\frac{\sigma + 2 \pi G R_{0}^2 \rho^2 + G R_0 \rho \sqrt{3 \pi \left(\pi R_{0}^2 \rho^2 + \frac{2\sigma}{G} \right)}  }
{\pi G R_{0}^2 \rho^2 - \sigma}
.
\end{equation}

Hence, in order for a binary to remain bound, both

\noindent{}$k_{1} \ge 2 + \sqrt{3} \approx 3.73$ (the strengthless limit)
and $R_0 > \sqrt{\sigma/(\pi G \rho^2)}$.  

\begin{figure}
\includegraphics[width=9cm]{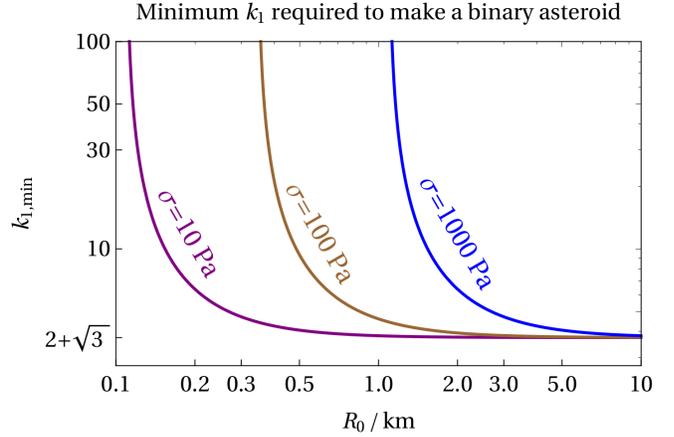}
\caption{
Minimum ratio of child asteroid radii ($k_1$; equation \ref{kratio}) for which a parent can break
up into a bound binary, for three different strengths. For strengthless asteroids, $k_{1,{\rm min}} = 2+\sqrt{3} \approx 3.73$
for all radii.
}
\label{kiminplot}
\end{figure}

Consequently, a strengthless parent asteroid of any
size can break apart into two bound components. For an asteroid with nonzero internal strength,
the minimum parent radius that could produce a binary asteroid is 110m, 350m and 1.1 km
for respectively $\sigma = 10, 10^2, 10^3$ Pa. For these asteroids, 
we plot $k_{1,{\rm min}}$ as a function of $R_0$ and $\sigma$ in Fig. \ref{kiminplot}.

Figure \ref{kiminplot}, combined with the instances of first fission from Figs. \ref{a5om0plots}-\ref{a200plot}, place
restrictions on where binary asteroids may be formed from YORP-induced break-up along the 
giant branch phases of stellar evolution. The times of first fission are independent of fragmentation prescription.

Strengthless rubble pile asteroids could form binary children for any parent radii where fission occurs: for example,
for radii of several km at distances of a few au, 1 km at a couple tens of au, or 0.1 km at a couple hundreds of au.
In contrast, the asteroids with the greatest internal strength, at $\sigma = 10^3$ Pa, are more restricted.
For them, at a distance of a 5 au, there is only a narrow radii range of 1-5 km where binary formation is possible.
At 30 au, the radii range is even narrower, between about 0.4-0.5 km. These results reinforce the notion that 
binary formation is more likely to occur at lower internal strengths. Nevertheless, if, like in the solar system, most asteroids
in post-main-sequence planetary systems are strengthless rubble piles, then we might expect binary asteroid formation to be 
ubiquitous.

\section{Discussion}

This large population of binary asteroids, however, may be short-lived. A bound binary asteroid
created from YORP-induced fission will then be subject to the BYORP, or binary YORP, effect
\citep{cukbur2005,cuk2007,golsar2009,waljac2015}. Unlike the YORP effect, the BYORP effect
alters the binary orbit. Both the semimajor axis and eccentricity of the orbit vary secularly \citep{mcmsch2010a}
and their rates of change are linearly proportional to the stellar luminosity \citep{mcmsch2010b}.

In addition to BYORP, another driver of evolution which must be treated self-consistently 
is the mutual tidal interaction between the child asteroids.
By assuming a constant Love number and constant stellar luminosity, \cite{jacsch2011} illustrated that when BYORP 
acts to shrink the semimajor axis and tides act to increase the semimajor axis, then the system could evolve towards an equilibrium state. 
Such a state provides a protection mechanism against destruction or escape, and hence allows the binary asteroid to be long-lived.

Determining whether equilibrium states could be maintained throughout and until the end of the giant branch phases of evolution 
would require detailed modelling. The stellar luminosity changes imply that the duration of an equilibrium state  
would be a function of when that state is achieved. 

Perhaps the greater source of complication, however, is the effect of tides.
The non-constancy of the Love number has important implications for tidal evolution \citep{efrwil2009,efrmak2013,makefr2013,coretal2014,bouetal2016}. Even for a simple Maxwell rheology and when tides from one of the bodies can be neglected, the semimajor axis and eccentricity vary non-monotonically \citep{veretal2019b}. This behaviour would be complicated further by the addition of a force (such as BYORP) which couples to both the orbital and spin tidal equations of motion, as does, by analogy, the Lorenz force in other contexts \citep{verwol2019}.

Regardless of the lifetime of binary asteroid systems, the total number formed would be a strong function of the size
of the monolithic constitutents (such as boulders or grains) of the rubble pile asteroids. We can obtain some realistic handle on 
the number of components of a rubble pile asteroid by considering Itokawa. \cite{micetal2008} provided size-frequency
statistics for the boulders on Itokawa, and determined a power law with an exponent of about $p = 2.8$.

Assume that a rubble pile asteroid of mass $M_{\rm asteroid}$ is composed of $N$ components ranging in mass from $M_{\rm min}$ to $M_{\rm max}$ which follow the following power-law distribution

\begin{equation}
\frac{dN}{dM} = u M^{-p}
.
\end{equation}

\noindent{Then}

\begin{equation}
u = \frac{N \left(1 - p\right)}
               {\left[M_{\rm max}^{1-p} - M_{\rm min}^{1-p}  \right]}
\end{equation}

\noindent{}and

\begin{equation}
M_{\rm asteroid} = \frac{u}{2-p} \left( M_{\rm max}^{2-p} - M_{\rm min}^{2-p} \right)
.
\end{equation}

\noindent{}Hence, we can solve for the number of components as

\begin{equation}
N = M_{\rm asteroid}\left( \frac{2-p}{1-p} \right) 
\left[ 
\frac
{M_{\rm max}^{1-p} - M_{\rm min}^{1-p}}
{M_{\rm max}^{2-p} - M_{\rm min}^{2-p}}  
\right]
.
\label{CompNum}
\end{equation}

If we assume $R_{\rm min} = 5$m and $R_{\rm max} = 10$m, then a density of $2$ g/cm$^3$ gives
$N \approx 5 \times 10^{-7} M_{\rm asteroid} /$kg.  In this case, an exo-Itokawa rubble pile would be composed of about
17,500 monolithic constituents\footnote{The total number of constituents is highly sensitive to $R_{\rm min}$. If
we neglect the $M_{\rm max}$ terms in equation \ref{CompNum}, then $N \propto R_{\rm min}^{-3}$.}. This value then would provide a constraint on the maximum allowed number of fissions for a
given prescription for the number of child asteroids produced (per fission) as well as their relative masses.

\section{Summary}

Almost every known white dwarf planetary system features asteroidal debris, highlighting the critical
need to understand how this debris is created during the giant branch phases of evolution.
This investigation is only the second dedicated work after \cite{veretal2014c} (Paper I) to analyze
aspects of the debris generated from giant branch YORP-based rotational fission, a process violent enough
to pulverize entire Main Belt analogues in extrasolar systems. 

Here we significantly expanded Paper I by considering multiple generations of fissions,
incorporating internal strengths and demonstrating conditions for binary asteroid formation,
all based on the analytical formalism of \cite{scheeres2018}.
Our key results are 

\begin{enumerate}

\item The final spin state is a weak function of stellar mass (Fig. \ref{steldep}),

\item Realistic nonzero internal strengths of rubble piles (up to 1 kPa) insufficiently protect
asteroids against giant branch YORP-induced rotational fission (Fig. \ref{sigmacrit}), 

\item Successive fissions occur in progressively smaller time intervals as the star ascends the giant branches,
despite the progressively faster failure spin rate (Fig. \ref{spinevo}), 

\item In most cases, there are either zero or more than ten fission events 
(or until the asteroid is broken down into its constituent boulders
and grains); between one and ten
fission events occur in only a narrow range of parameter space (Figs. \ref{a5om0plots}-\ref{a200plot}), 

\item Binary asteroid formation from strengthless rubble piles is easy to generate and ubiquitous along the giant branch phases 
(Fig. \ref{kiminplot}), but the sustainability of that configuration until the white dwarf phase is in question (Section 3).

\end{enumerate}

\section*{Acknowledgements}

{\rev We thank the referee for helpful comments which have improved the manuscript.} The authors also acknowledge initial discussions for this project at the 9th Workshop on Catastrophic Disruptions in the Solar System in Kobe. DV gratefully acknowledges the support of the STFC via an Ernest Rutherford Fellowship (grant ST/P003850/1).

\label{lastpage}
\end{document}